\documentclass[aps,prb,preprint,amssymb,amsfonts]{revtex4-1} 

\usepackage{amsthm,graphicx,epsfig}

\newcommand{\nf}{F_{\rm N}}

\begin{document}

\title[]{Comparison of the lateral retention forces on sessile and pendant 
water drops on a solid surface}

\author{Rafael de la Madrid}
\email{rafael.delamadrid@lamar.edu}

\author{Taylor Whitehead}
\author{George M.~Irwin}
\affiliation{Department of Physics, Lamar University,
Beaumont, TX 77710}

\date{January 20, 2015}

\begin{abstract}
\noindent We present a simple experiment that demonstrates how a water 
drop hanging 
from a Plexiglas surface (pendant drop) experiences a lateral retention 
force that is comparable to, and in some cases larger than, the lateral 
retention force on a drop resting on top of the surface (sessile drop). The 
experiment also affords a simple demonstration of the Coriolis effect in two 
dimensions.
\end{abstract}

\maketitle

\section{Introduction}
\label{sec:intro}

When two solid surfaces are in contact and we try to set one of the
surfaces in motion relative to the other, there appears
a static frictional force $\vec{f}_{\rm s}$ that
opposes the motion. For rough 
surfaces, it 
is found experimentally that the static 
frictional force increases up to a maximum value 
$\vec{f}_{\rm s}^{\rm max}$, and if the applied force
is greater than this maximum value, then the surfaces will start 
to move relative to each other. Once the surfaces are in relative motion,
there is a kinetic frictional force $\vec{f}_{\rm k}$ that 
opposes the motion. When the surfaces are rough, $\vec{f}_{\rm s}^{\rm max}$ and
$\vec{f}_{\rm k}$ are usually proportional to the normal force 
(the load).\cite{ZEMANSKY,HRW,PALMER1,PALMER2,PALMER3,MANEY,BOWDEN,KRIM1,KRIM2,ROBBINS} 

When a liquid drop is in contact with a solid surface and we try 
to slide the drop relative to the solid surface, there appears a
lateral retention force (due to adhesion) that opposes the motion
and can be thought of as the liquid-solid analog of solid-solid friction. An 
everyday example of such a force
can be observed when water drops get stuck on the interior surface of a drinking
glass. Due to surface tension, pressure, and the 
deformability of liquid drops, liquid-solid retention forces
exhibit effects that do not appear in solid-solid friction. One 
such effect was presented by Tadmor {\it et al.},~\cite{TADMORPRL} who
found experimentally that the lateral retention force on hexadecane drops 
is larger when the drops hang from the surface compared to when they
rest on top of the surface. To demonstrate this effect, these researchers 
placed hexadecane drops either on top (a sessile drop) or on the bottom 
(a pendant drop) of a rotating platform. The 
platform was then rotated at an increasing angular speed until the resulting 
centrifugal force caused the drops to slide. Naively, in analogy with friction, one might think that a
sessile drop would be more ``stuck'' to the surface than a pendant drop, and 
therefore a pendant drop would be easier to slide.
However, the opposite turns out to be true: the bottom (pendant) drop is in fact
\textit{harder} to slide. It would seem then that the retention force on the 
bottom (pendant) drop is larger than the retention force on the top 
(sessile) drop. 

While the experiment of Ref.~\onlinecite{TADMORPRL} was performed using a
sophisticated (and expensive) apparatus called a 
\textit{centrifugal adhesion balance}, the purpose of 
the present paper is to introduce a straightforward and inexpensive 
experiment that demonstrates the same effect. In this experiment,
one can see with the naked eye that the force required to slide a
water drop hanging from a Plexiglas sheet is comparable to, and 
sometimes larger than, the force required
to slide a drop resting on the sheet. In addition, the differing tracks 
that the drops leave as they slide on the sheet also show the same 
result.

The counter-intuitive nature of the experiment can be better appreciated
by comparison with a similar experiment performed with magnets, where
the role of the adhesive force is played by the magnetic force. If we
place a stack of magnets on top and on the bottom of a rotating steel bar, we
would expect that the magnets on top will be more ``stuck'' to the bar than
the ones on the bottom. This is because the normal force (and hence the
frictional force) on the magnets on top are larger than on those on the
bottom. Thus, the magnets on the bottom should be easier to slide than the
magnets on top.

In Sec.~\ref{sec:magnets}, we carry out the experiment with magnets
and observe that the bottom magnets are always 
{\it easier} to slide than the top magnets. In Sec.~\ref{sec:sdrops}, 
we carry out the experiment using water drops and observe that
the bottom drop is on average {\it harder} to slide than the top 
drop. In Sec.~\ref{sec:theory}, we 
provide a qualitative description to explain
the results of the experiment, and Sec.~\ref{sec:conclusions} contains 
our conclusions.

\section{The magnets}
\label{sec:magnets}

\begin{figure}[ht!]
\includegraphics[width=8.5cm]{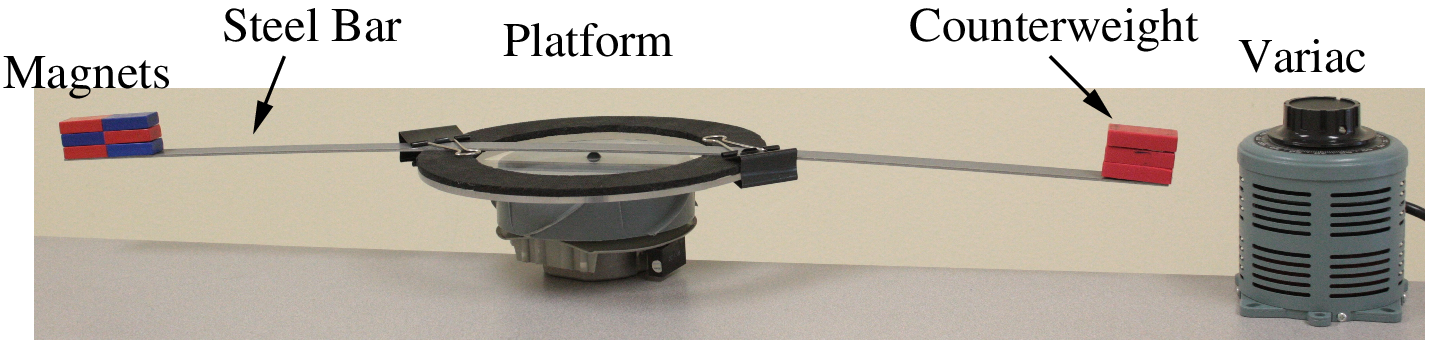}
\caption{Apparatus used for the magnet experiment. The fan motor
is under the platform.}
\label{FIG:steelbar}
\end{figure}

Figure~\ref{FIG:steelbar} shows our experimental apparatus. A rotating 
platform is mounted on top of a motor.\cite{FN0} A 
variable transformer (a Variac) provides speed control for 
the motor that makes the platform rotate. A steel bar is attached to 
the rotating platform with two binder clips. A stack of three bar magnets 
is placed at one end of the steel bar, first on top and then on the bottom. In 
order to keep the platform balanced, three additional magnets are placed on 
the other end of the steel bar as a counterweight. These additional magnets 
are strong enough so that they remain fixed throughout the experiment.

As viewed from a (non-inertial)
reference frame that rotates with the platform, there are four horizontal
forces acting on the magnets.\cite{CASSIDAY} The first force is the 
centrifugal force $\vec{F}_{\rm cf}$, which increases with angular speed and 
is responsible for ultimately making the magnets move.
The second force is the frictional force $\vec{f}$: $\vec{f}_{\rm s}$
when the magnets do not move relative
to the steel bar, and $\vec{f}_{\rm k}$ once the magnets start to move.
The third force is the \textit{transverse} force $\vec{F}_{\rm tr}$
(also known as the \textit{azimuthal} or \textit{Euler} force),
which acts only when there is an angular acceleration.
Lastly, there is the Coriolis force $\vec{F}_{\rm co}$, which acts 
only when the magnets are moving relative to the steel bar. 
In addition, there are three forces acting on the magnets in the
vertical direction: their weight $\vec{W}$, the normal 
force $\vec{F}_{\rm N}$, and the magnetic attraction $\vec{F}_{\rm a}$ between 
the steel bar and the magnets.

In our experiment, once the magnets begin to slide on the steel bar, they 
quickly lose contact with the bar and fly off. Therefore, we only need to 
describe the forces that act on 
the magnets when they are at rest relative to the steel bar. In such a 
situation, the Coriolis force is zero. In addition, at the moment the magnets
are about to slide off, the 
transverse force can be neglected compared to 
the centrifugal force (see Appendix~\ref{sec:transf}).
Furthermore, because this is the first instant when the static frictional force is
overcome, we can safely neglect the transverse force for the entire problem. 

\begin{figure}[ht!]
\includegraphics[width=8.5cm]{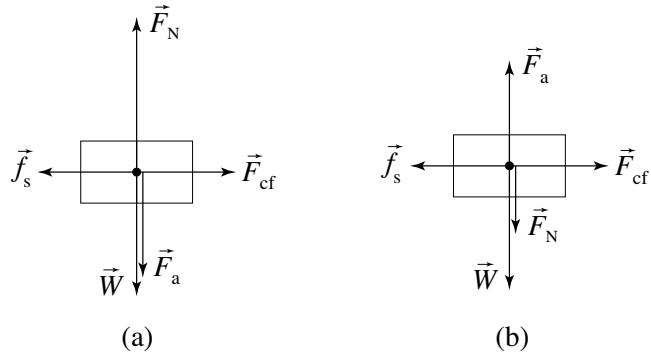}
\caption{Free-body diagram of the magnets when they are placed (a) on top
and (b) on the bottom of the steel bar. The transverse force $\vec{F}_{\rm tr}$ 
(not shown in the figure) is perpendicular to the page. The centrifugal
force is balanced by the static frictional force until the magnets
start sliding.}
\label{FIG:free-body}
\end{figure}

Figure~\ref{FIG:free-body} shows a free-body diagram of the 
magnets when they are at rest relative to the steel bar. In
the horizontal direction, Newton's second law yields
\begin{equation}
      f_{\rm s}=F_{\rm cf}= mr\Omega ^2 \, ,
           \label{feqfc}
\end{equation}
where $m$ is the mass of the magnets, $r$ is the distance from
the axis of rotation to the magnet's center-of-mass,
and $\Omega$ is the angular speed of 
rotation. Meanwhile, in the vertical direction we have
\begin{equation}
        \nf = F_{\rm a} \pm mg,
                 \label{nf}
\end{equation}
where the plus sign holds when the magnets are on top of the bar
and the minus sign holds when the magnets are on the bottom.
Thus, the maximum static frictional force on the magnets is given by
\begin{equation}
      f_{\rm s}^{\rm max} = \mu_{\rm s} (F_{\rm a} \pm mg),
                 \label{nffm}
 \end{equation}
where $\mu _{\rm s}$ is the coefficient of static friction. Because the 
static frictional force balances the centrifugal force
[see Eq.~(\ref{feqfc})], we see that as the angular speed increases,
the frictional force will increase until it reaches its maximum value.
Denoting by $\Omega _{\rm c}$ the angular speed 
at which the centrifugal force is equal to the maximum (static) 
frictional force, Eqs.~(\ref{feqfc})--(\ref{nffm}) then yield
\begin{equation}
      \Omega _{\rm c}^2 = \frac{\mu_{\rm s}}{mr} (F_{\rm a} \pm mg).
                 \label{nffmo}
\end{equation}
As expected, this equation shows that the angular speed at which the 
magnets fly off is lower when the magnets are placed on the bottom than 
when they are placed on top. Hence, for a platform that accelerates at 
a constant angular acceleration, the magnets on the bottom should fly 
off before the magnets on top.

The magnets used in our experiments come inside a plastic case, and the 
coefficient of static friction between the plastic case and the steel bar was
measured by tilting the steel bar with the case on top. By measuring
the critical angle $\theta _{\rm c}$ at which the case begins to slide, 
the well-known relation $\mu_{\rm s}=\tan \theta _{\rm c}$ yields the coefficient
of static friction. In our case, we measured $\mu_{\rm s}= 0.28 \pm 0.03$. We also 
measured the magnetic attraction between the magnets and the steel bar by
placing the magnets on the bottom of the steel bar and adding mass to the magnets
until they lost contact with the bar. The total weight (magnets plus
added mass) yielded a magnetic attraction of 
$F_{\rm a}= 2.078 \pm 0.001$\,N.

The mass of the magnets (including the plastic
case) was measured to be $m=0.141\pm 0.001$\,kg and they were placed
a distance $r=0.24\pm 0.01$\,m from the axis of rotation.
These values were used to calculate a theoretical critical
angular speed $\Omega_{\rm c}^{\rm (th)}$ using Eq.~(\ref{nffmo}).
The experimental critical angular speed $\Omega_{\rm c}^{\rm (exp)}$ was also measured
using PASCO's smart timer ME-8930. As can be seen in Table~\ref{t:m}, there is
a reasonable agreement between the theoretical and experimental values.

\begin{table}[h!]
\caption{Theoretical and experimental critical angular speeds for magnets (in rad/s).}
\vspace{6pt}
\begin{tabular}{lcc}
\hline\hline
     &$\Omega _{\rm c}^{\rm (th)}$
     &$\Omega _{\rm c}^{\rm (exp)}$ \\
\hline
   Top & $5.4 \pm 0.3$  & $5.7 \pm 0.5$ \\
   Bottom\quad \ & $2.5 \pm 0.1$    & $2.7\pm 0.5$  \\ 
\hline\hline
\end{tabular}
\label{t:m}
\end{table}

We also performed a second experiment with magnets that resembles the
experiment with drops more closely. We placed two equal stacks of magnets
on the top and bottom of the the steel bar at the same time. Then we turned
on the motor and increased the angular speed until the magnets flew off the
steel bar. In the 20 times that we performed this second experiment, the stack
on the bottom flew off before the stack on top 100\% of the time, as expected.

\section{The sliding drops}
\label{sec:sdrops}

\subsection{Water drops on Plexiglas}

\begin{figure}[ht!]
\includegraphics[width=8.5cm]{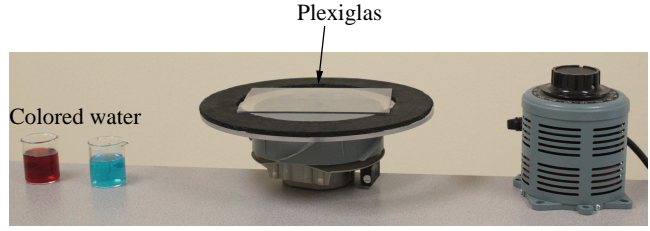}
\caption{Apparatus used for the drop experiment. A Plexiglas
sheet is placed on top of the platform. A blue drop 
is placed on the top of the Plexiglas sheet and a red drop 
is placed on the bottom. When the angular speed 
of the platform reaches a critical value, the drops 
begin sliding on the Plexiglas sheet.}
\label{FIG:platform}
\end{figure}

In the third part of the experiment, we replaced the steel bar with
a Plexiglas sheet (see Fig.~\ref{FIG:platform}). We poured water into 
two beakers and added 
red and blue food coloring to the water. We then placed a red drop on 
one side of the Plexiglas sheet,
gently flipped the sheet, and put a blue
drop on the other side. Thus, when we put the Plexiglas sheet on the
rotating platform, the blue drop was on top and the red drop was on 
the bottom. Both drops were placed the same distance from the
axis of rotation, so the centrifugal force acting on them was always the
same. To prevent the Plexiglas sheet from flying off the platform, it
was attached with masking tape.

Similar to the magnet experiment, when the angular speed reached 
a critical value, the drops started to slide on the Plexiglas 
surface. However, contrary to what happened with the magnets, the drop on 
top of the Plexiglas sheet started to slide (on average) before the drop on 
the bottom. For 10-$\mu$L drops, we found that the drop on top started to 
slide first in about 60\% of our runs, in contrast to the magnet experiment,
where the magnets on top never slid first.

\begin{figure}[ht!]
\includegraphics[width=8.5cm]{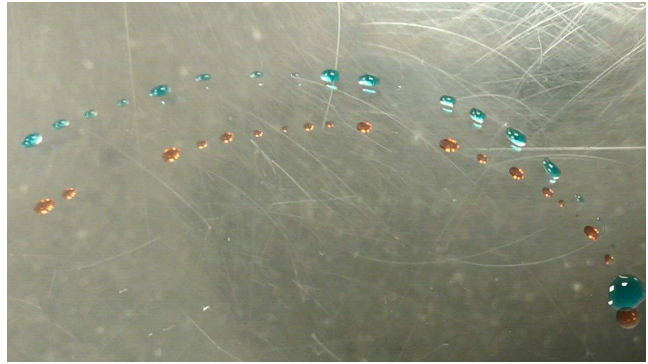}
\caption{Picture of the trails left by the water drops on the 
Plexiglas sheet. The upper (blue) droplets are left behind by the sessile
drop; the lower (red) droplets are left behind by the pendant drop. The 
trajectory of the pendant drop is 
more curved than the trajectory of the sessile drop.}
\label{FIG:trajectory}
\end{figure}

When the drops slide on the Plexiglas sheet, the Coriolis force deflects
their motion and the drops leave behind a trail of smaller droplets
that mark their trajectories.
Figure~\ref{FIG:trajectory} shows the trajectories of 
the red (pendant) and blue (sessile) drops.
It is clear from this figure that the sessile drop follows a 
trajectory that 
is less curved than the trajectory of the pendant drop. Thus, the 
Coriolis force acting on the sessile drop is smaller than the Coriolis 
force acting on the pendant drop. Because the Coriolis force
is increasing with time,
we conclude that the sessile drop 
started to slide earlier than the pendant drop.

As can be seen in Fig.~\ref{FIG:trajectory}, because our 
platform rotates clockwise, the drops in our experiment deflect to the left
due to the Coriolis force. One can easily reverse the direction of rotation to
produce a deflection to the right.\cite{VIDEO1} Thus, this experiment 
also affords a simple demonstration of the Coriolis effect.

\subsection{Troubleshooting}
\label{sec:troubleshooting}

Flipping the Plexiglas sheet to put a drop on the bottom is
the most delicate step of the experiment. If the flip is done too slowly, the drop
may begin sliding on the surface, and it will take a much smaller 
centrifugal force to get it sliding again. Hence, when you flip the sheet,
it is crucial to do it quickly and to use an axis of rotation that goes through the drop,
so that the drop is anchored in place.

Another delicate step in the experiment is to place the top and bottom 
drops at the same distance from the axis of rotation so that
the centrifugal force acting on both drops is the same.
Fortunately, it is straightforward to see if these steps have been
carried out correctly. If the contact areas of the the drops are the same, then
the bottom drop has not slid significantly, and both drops will be at the same
distance from the axis of rotation.

In some runs, both drops start to slide at about the same
time, and it may be difficult to visually determine which drop started 
to slide first. Thus, it is useful to mount a video camera on the rotating
platform. For example, a cell-phone or a GoPro\cite{GoPro} camera
(both of which are very light) can be mounted directly to the
platform using tape to get a surface view of the motion of the drops.
Alternatively, an elevated view can be obtained by mounting the camera on
top of a piece of foam or using the small mount that comes with a GoPro
camera. In our experiments, we used a GoPro Hero3 camera, which allows wireless
transmission of the video so you can view the movie without having to unmount
the camera. The resulting video footage~\cite{VIDEO2} shows how (on average) 
the sessile (top) drop starts to slide first.

The size of the drops is an important factor that affects the 
critical angular speed at which the drops begin to slide. When two drops
are placed on the same side of the Plexiglas sheet (say on top)
at the same distance from the axis of rotation, the
larger drop starts to slide before the smaller one. Thus, a 
micropipette should be used so that each drop
has the same volume.

When one places a drop on the sheet, it is important that the tip of the
micropipette is not touching the sheet. Otherwise, when you remove the tip it is
easy to inadvertently move the drop. Conversely, the tip should not be 
too far from the sheet so that when the drop falls it does not splash.

Let us denote by $t_{\rm B}$, $\Omega _{\rm B}$, and $f_{\rm B}$ the time, angular
velocity, and centrifugal force at which the bottom (pendant) drop 
begins sliding. Then
\begin{equation}
         f_{\rm B}=mr\Omega _{\rm B}^2 = \rho V r\Omega _{\rm B}^2
                \label{bf}
\end{equation}
and
\begin{equation}
        \Omega _{\rm B}= \alpha t_{\rm B} ,
                \label{Ob} 
\end{equation}
where $m$, $\rho$, $V$, $r$, and $\alpha$ are, respectively, the mass of the
drop, the density of the liquid, the volume of the drop, the distance from 
the axis of rotation, and
the angular acceleration of the platform. By combining Eqs.~(\ref{bf})
and~(\ref{Ob}), we obtain
\begin{equation}
             t_{\rm B}=\frac{1}{\sqrt{\rho Vr\alpha ^2}} \sqrt{f_{\rm B}}  ,
             \label{tbottom}
\end{equation}
with a similar equation (for $t_{\rm T}$) holding for the top drop. We
then find that the time difference $\Delta t$ between when the two
drops begin sliding is given by
\begin{equation}
         \Delta t = t_{\rm B}-t_{\rm T}=\frac{1}{\sqrt{\rho Vr\alpha ^2}} 
              \left(\sqrt{f_{\rm B}} - \sqrt{f_{\rm T}}\right).
          \label{tdiffe}
\end{equation}

Equation~(\ref{tdiffe}) shows that in order to increase $\Delta t$, one must
decrease either the volume of the drops, the distance from the axis of
rotation, or the angular acceleration (or some combination of these quantities).
In our  experiments, the best results were obtained with
$V\approx 10\,\mu$L, $r\approx 5$\,cm, and $\alpha \approx 1.5$\,rad/s$^2$.
Of course, one could decrease the volume of the drops further to obtain 
an even larger $\Delta t$, but it becomes difficult to visually
follow the motion of such small drops so a camera would be necessary.

When we dried the Plexiglas sheet after each run, we sometimes inadvertently
created static electricity. In order to prevent static 
electricity from building up on the sheet, we found it useful to be barefoot 
while drying the sheet. We also found it useful to rub the sheet
on a grounded metal pole.

We have also performed these experiments with other surfaces
(Lexan, glass, and PVC) and different water-based liquids
(vinegar and wine). In all of these other experiments, we did not
observe the effect, either because it is not present or because
it is too small to be seen with our apparatus. This seems to indicate
that the effect is not universal and depends on the liquid/solid 
combination. This would not be terribly surprising, given that the adhesive
force depends on both the liquid and the solid. However, it is still
an open question whether only some liquid/solid combinations exhibit this effect.

\subsection{Systematics}
\label{sec:systematics}

In order to eliminate some obvious systematic effects as the source of 
our results,
we varied the experiment in several ways. First, we flipped the Plexiglas 
sheet, so what was the bottom face became the top face, and vice versa. Second,
we varied whether the red or blue drop was placed on the top or bottom of the sheet.
Third, we placed the drops at different locations on the 
sheet, and used both distilled and tap water. In addition, we used 
six different Plexiglas surfaces to perform the experiment. Some
surfaces appeared very smooth, and others appeared a bit 
scratched. In all the possible variations of the experiment, we observed 
that, with 10-$\mu$L drops, the top (sessile) drop started 
to slide first in about 60\% of the runs, whereas the bottom (pendant) drop
was first in about 37\% of the runs (in about 3\% of the runs, both drops
started to slide at about the same time). With 20-$\mu$L drops, the
top (sessile) drop started to slide first in about 40\% of the runs, whereas 
the bottom (pendant) drop was first in about 47\% of the runs 
(in about 3\% of the runs, both drops started to slide at about the same 
time).\cite{NOTE} This 
result is in sharp contrast with the magnet experiment, where the top 
magnet never slid first.

We didn't treat the Plexiglas surfaces in any way, and therefore the roughness
of such surfaces is fairly inhomogeneous, that is, some spots produce
a larger retention force than others. In the ideal case that the surfaces were
perfectly homogeneous, every spot would produce the same retention force,
and we would obtain the same result on every run of the experiment. If we 
identify the average result of our experiment with the result of the ideal
case, we would have that on an ideal surface, 10-$\mu$L pendant drops would 
always slide slightly before than 10-$\mu$L sessile drops, and hence the
lateral retention force on 10-$\mu$L pendant drops would be slightly larger than
on sessile drops. For 20-$\mu$L drops, we would have that 
the lateral retention force on pendant drops is slightly smaller than,
though comparable to, such force on sessile drops.

It is a common experience that when stains are cleaned quickly, they are easier
to remove than when they are left to sit for a while. The same happens
to liquid drops sitting
on a surface. As discussed in Ref.~\onlinecite{TADMORPRL}, the
lateral force needed to move a liquid drop increases with the time that
the drop sits on the surface. One may therefore wonder if the 
order in which the drops are placed in our experiment has any effect 
on the results. In order to check this, we performed several runs of 
the experiment where we first placed the top drop, then flipped the sheet
and placed the bottom drop, and then flipped the sheet again.
In this way, the top drop was in contact with
the Plexiglas sheet a few seconds longer than the bottom drop. For
these trials, we also observed that on average the top drop started to
slide before the bottom drop.

In about 10\% of the runs, we observed that the trail left by the
top drop was more curved than the trail left by the bottom drop, even though
the top drop started to slide first. This puzzling result has been interpreted
in the following way. As the drops slide on the sheet, the rate at which
they lose mass by leaving droplets behind is not exactly the same,
presumably because the roughness of the Plexiglas sheet is not
homogeneous. If the 
bottom drop loses mass at a higher rate than the top drop, the Coriolis
force on the bottom drop may become smaller than the Coriolis force on
the top drop, and the bottom drop may 
leave a trail that is
less curved than the trail of the top drop, even though the top
drop began sliding first.

One of the main concerns of any experimental study of drops sliding
on substrates is the presence of contaminants. In our experiments, we 
made little effort to remove contaminants from the Plexiglas surfaces, water, 
or food coloring. We limited ourselves to cleaning the surfaces with 
ethyl alcohol and paper towels. Because possible contaminants did not 
seem to alter the result of the experiment, we believe that the experiment can
be easily reproduced.

\section{A tentative theoretical explanation}
\label{sec:theory}

Although droplets in contact with solid 
surfaces have been extensively
studied,\cite{DEGENNES,ERBIL,BORMASHENKO,DEGENNESRMP,BONNRMP}
the behavior of a droplet just before it begins to move is still not well 
understood. In particular, there is no accepted theory that explains why
a drop hanging from a surface experiences a larger lateral retention
force than a drop resting on the surface. Nevertheless, in this section 
we would like to present some theoretical attempts to explain such
experimental results. Our presentation is approximate, and the reader
is referred to the literature for more
detailed explanations. In particular, we will omit thermodynamical and 
energy balance arguments,
even though they are more fundamental than force balance 
arguments.\cite{PELLICER}

\subsection{Drop resting on a horizontal surface: The Young equation}

\begin{figure}[ht!]
\includegraphics[width=8.5cm]{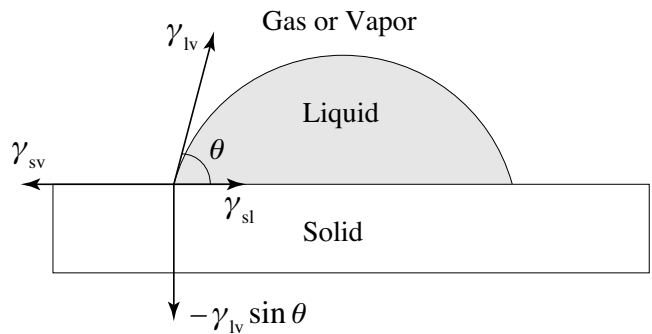}
\caption{Surface tensions at the three-point contact line of a drop resting on
a surface.}
\label{FIG:youngequation}
\end{figure}

\begin{figure}[ht!]
\includegraphics[width=8.5cm]{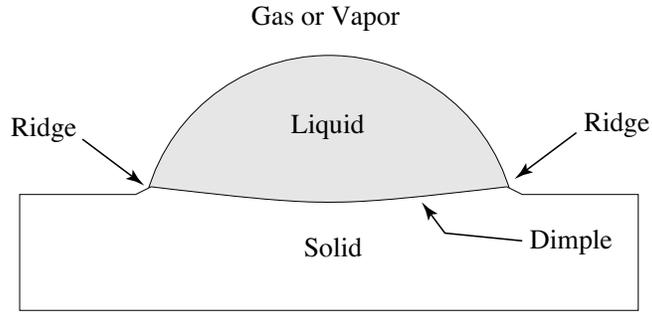}
\caption{Elastic deformation (not drawn to scale) of the solid substrate due to the vertical
component of the liquid-gas surface tension (ridge) and to the Laplace 
pressure (dimple).}
\label{FIG:deform}
\end{figure}

A liquid drop resting on a solid surface acquires a shape that minimizes
the sum of its surface and gravitational potential energies
(see Fig.~\ref{FIG:youngequation}). The contact angle $\theta$ is defined
as the angle formed by the liquid-vapor and the liquid-solid
interfaces. The line where the solid, liquid, and vapor phases co-exist
is called the three-phase contact line. The three-phase contact line
is also the contour line of the nominal area of contact between the liquid
and the solid. When the drop is at 
rest, the sum of the tensions at each point of the contact line must
be equal to zero. From Fig.~\ref{FIG:youngequation}, the equilibrium of 
surface tensions in the direction parallel to the solid-liquid interface leads
to the Young equation
\begin{equation}
        \gamma _{\rm sv}= \gamma _{\rm sl} + \gamma _{\rm lv} \cos \theta \, ,
          \label{youneq}
\end{equation}
where $\gamma _{\rm sv}$, $\gamma _{\rm sl}$, and $\gamma _{\rm lv}$ are the
solid-vapor, solid-liquid, and liquid-vapor surface 
tensions, respectively. Figure~\ref{FIG:youngequation} also shows that the 
liquid-vapor surface tension has a component that is perpendicular to the 
solid-liquid interface. Such a component will pull on the surface and deform 
it a small amount. In equilibrium, the vertical component is balanced with a 
downward tension of magnitude $\gamma _{\rm lv} \sin \theta$ that arises from 
the strain on the surface of the solid. This strain causes the formation of 
a small ridge on the surface of the solid along the
contact line (see Fig.~\ref{FIG:deform}). For water
on Plexiglas, such a ridge is of the order of nanometers, but for
softer surfaces it can be big enough to produce measurable
effects.\cite{LESTER,RUSANOV,SHANAHAN,PERICET,ROMAN,LUBARDA,STYLE1,STYLE2}

The Laplace equation is another important equation that describes a
drop resting on a surface. The Laplace equation relates the pressures
inside ($p_{\rm in}$) and outside ($p_{\rm out}$) the drop 
with the shape of the drop:
\begin{equation}
        \Delta p = p_{\rm in}-p_{\rm out} =
                    \gamma_{\rm lv}\left( \frac{1}{R_1}+\frac{1}{R_2}\right) \, ,
\end{equation}
where $R_1$ and $R_2$ are the principal radii of curvature of
the surface of the drop. The Laplace pressure (in addition to
the weight of the drop) pushes down onto 
the surface over the contact area, creating a dimple. The combined effects of
the Laplace pressure and the vertical component of the surface tension
produce a crater-like deformation of the surface, as shown 
in Fig.~\ref{FIG:deform}.

\subsection{Drop resting on an incline and on a rotating platform}

\begin{figure}[ht!]
\includegraphics[width=8.5cm]{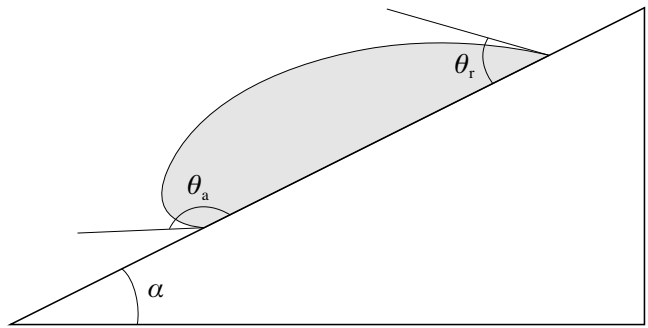}
\caption{Deformation of a drop on an incline.}
\label{FIG:incline}
\end{figure}

When a drop rests on an incline, the component of the drop's weight
parallel to the incline ($mg\sin \alpha$) will deform the drop (see 
Fig.~\ref{FIG:incline}), and the contact angle will change from its value
on a horizontal surface. The contact angles that the leading and trailing 
edges of the drop make
with the incline right before the drop starts to slide are called the
advancing $(\theta _{\rm a}$) and receding $(\theta _{\rm r}$)
contact angles. 

Because the drop is not sliding down the incline, there must be a force
that counterbalances the component of gravity parallel to the surface.
When the drop is about to slide, the force that keeps the drop from sliding 
down the incline is given 
by\cite{FRENKEL,FURMIDGE,DUSSAN,EXTRAND,KIM,ELSHERBINI,GARIMELLA} 
\begin{equation}
       f_{\rm max} = k w \gamma _{\rm lv} (\cos \theta _{\rm r}-\cos \theta _{\rm a})
             \, ,
       \label{fpw}
\end{equation}
where $w$ is the width of the drop and $k$ is a dimensionless
quantity that depends on several factors, including the shape of the 
triple line and the shape of the drop. 

Intuitively, one can understand the origin of Eq.~(\ref{fpw}) as 
follows. When the drop is at rest on a horizontal surface, the forces per 
unit length parallel to the solid surface that act on a given
point of the contact line are $\gamma _{\rm sv}$, $\gamma _{\rm sl}$, and
$\gamma _{\rm lv} \cos \theta$. By the Young equation (\ref{youneq}), at every point 
on the contact line, the pull produced by $\gamma _{\rm sv} - \gamma _{\rm sl}$ 
is compensated by the pull provided by $\gamma _{\rm lv}\cos \theta$. Because
the net force per unit length at each point of the contact line is zero,
the contact line does not move.

We can look at the equilibrium of the contact line in a slightly different
way. Let us assume that the contact line is in the shape of a circle. On each
point of the contact line, we have two tensions acting
on one such point, $\gamma _{\rm sv} - \gamma _{\rm sl}$ and
$\gamma _{\rm lv}\cos \theta$. Let us now consider two points $P_1$ and
$P_2$ on the contact line that are symmetrically placed 
with respect to the center of the drop. Then, by symmetry, we have that the
pull on $P_1$ due to $\gamma _{\rm sv} - \gamma _{\rm sl}$ is exactly canceled by
the pull on $P_2$ due to $\gamma _{\rm sv} - \gamma _{\rm sl}$. Similarly,
the pull on $P_1$ due to $\gamma _{\rm lv}\cos \theta$ is exactly canceled
by the pull on $P_2$ due to $\gamma _{\rm lv}\cos \theta$. Thus, when we add the
pulls due to $\gamma _{\rm sv} - \gamma _{\rm sl}$ along the contact line, they 
add up to zero. Similarly, when we add the pulls due to 
$\gamma _{\rm lv}\cos \theta$ along the contact line, they also add up to zero. 

When we place a drop on an incline, gravity deforms the drop, and the
pulls due to surface tensions do not cancel any more. Let us consider our
point $P_1$ to be the advancing edge of the drop, and our point 
$P_2$ the receding edge. In the case of a drop on an incline, the pull due 
to $\gamma _{\rm sv} - \gamma _{\rm sl}$ on the advancing edge
{\it is} canceled by the pull due to $\gamma _{\rm sv} - \gamma _{\rm sl}$ 
on the receding edge. However, the pull due to 
$\gamma _{\rm lv}\cos \theta _{\rm a}$ on the advancing edge is {\it not} 
canceled by the pull due to $\gamma _{\rm lv}\cos \theta _{\rm r}$ on the 
receding edge. The net pull
per unit of length on the advancing and receding edges of the drop is then 
$-\gamma _{\rm lv} (\cos \theta _{\rm r}-\cos \theta _{\rm a})$. If the drop
is not moving, there must exist a retention force per unit length
$\gamma _{\rm lv} (\cos \theta _{\rm r}-\cos \theta _{\rm a})$
that cancels this pull. If we assume that the drop is symmetric with respect to 
an axis that is parallel to the direction of propagation, and if we divide
the drop into an ``advancing half'' and a ``receding half,'' we can apply the
above analysis to any symmetrically-placed pair of points, with $P_1$ in the
``advancing half,'' and $P_2$ in the ``receding half.'' The retention force 
per unit length on these symmetrically placed points is also 
$\gamma _{\rm lv} (\cos \theta _{\rm r}^*-\cos \theta _{\rm a}^*)$, where
now $\theta _{\rm a}^*$ and $\theta _{\rm r}^*$ are the contact angles
at $P_1$ and $P_2$. By assuming (a)
that $\theta _{\rm a}^*$ remains constant and equal to 
$\theta _{\rm a}$ over the ``advancing half'' of the drop, (b) that
$\theta _{\rm r}^*$ remains constant and equal to 
$\theta _{\rm r}$ over the ``receding half'' of the drop, and (c)
that the triple contact line is a perfect circle
(or another convenient but unrealistic line~\cite{DUSSAN}), and by integrating
the retention force per unit length over the length of the triple contact 
line, one arrives at an expression for the
retention force given by Eq.~(\ref{fpw}), with $w$ equal to the diameter
of the drop and $k=1$.\cite{DUSSAN} A fudge factor $k$ is then introduced 
in Eq.~(\ref{fpw}) to account for the fact that the triple contact
line is not a perfect circle, and that the contact angle does not 
remain constant over the triple contact line.

When the drop is placed on top of a rotating platform, it also gets deformed
by the centrifugal force. Since inertial forces are equivalent to 
gravitational forces, the deformation of a drop on a rotating platform
is similar to its deformation on an incline. In particular, 
Eq.~(\ref{fpw})
is also assumed to yield the maximum lateral retention force on a drop that is
placed on a rotating platform.~\cite{EXTRAND}

When a drop is placed on the bottom of a rotating platform, Eq.~(\ref{fpw})
is also used to calculate the retention force, although for a pendant drop the
values of $\theta _{\rm r}$, $\theta _{\rm a}$, $k$, and $w$ are in general 
different from those of a sessile drop.

\subsection{Two possible explanations}

In Ref.~\onlinecite{TADMORPRL}, the advancing and receding contact
angles of the sessile and pendant drops were measured to be
$\theta _{\rm a,s}=35.5^\circ$, $\theta _{\rm r,s}=30.3^\circ$,
$\theta _{\rm a,p}=40.0^\circ$, and $\theta _{\rm r,p}=34.7^\circ$. For the size of
drops used in Ref.~\onlinecite{TADMORPRL}, surface tension dominates over
gravity, and the shape of the drops is nearly a spherical cap. Thus, the 
widths of the sessile and pendant drops are very similar and, although there 
was no attempt to determine it, the factor $k$ in Eq.~(\ref{fpw}) can be
assumed to be similar for sessile and pendant drops (because the drops are 
so similar in shape).\cite{PC} Thus, according to Eq.~(\ref{fpw}), the
retention forces for sessile and pendant drops are
$f_{\rm max}^{\rm sess}= kw\, 0.0493$ and $f_{\rm max}^{\rm pend}= kw\, 0.0561$. In
particular, Eq.~(\ref{fpw}) yields a larger retention force for
pendant drops than for sessile drops.

Although this seems like a plausible explanation, it has been
argued~\cite{TADMORPRL,TADMOR1,TADMOR2,TADMOR3} that Eq.~(\ref{fpw})
needs to be replaced by a different equation. The reason is that the 
experimental ratio of the retention forces on sessile and pendant drops in
the experiment of Ref.~\onlinecite{TADMORPRL} is
\begin{equation}
     \frac{f_{\rm max}^{\rm pend}}{f_{\rm max}^{\rm sess}}    = 1.27 
      \quad \text{(experimental),}
       \label{er}
\end{equation}
whereas Eq.~(\ref{fpw}), under the assumption that $k$ and $w$ are the
same for sessile and pendant drops, yields a ratio of 1.14.

As mentioned above, 
the vertical component of the liquid-vapor
tension and the Laplace pressure deform the solid surface and produce
a crater-like deformation. Although such deformation
is small, it has been proposed that it leads to a stronger interaction
between the molecules of the liquid and the 
solid.\cite{TADMORPRL,TADMOR1,TADMOR2,TADMOR3} By taking into account 
this stronger interaction, Eq.~(\ref{fpw}) is modified to\cite{TADMORPRL,TADMOR1,TADMOR2,TADMOR3}
\begin{equation}
       f_{\rm max} = \frac{4\gamma _{\rm lv}^2 \sin \theta}{E} 
              (\cos \theta _{\rm r}-\cos \theta _{\rm a})
             \, ,
       \label{fpwm}
\end{equation}
where $E$ is an interfacial modulus that is similar to the elastic modulus of 
a solid but that only takes into account the deformation of the outermost layer
of the solid (which is the only part of the solid in direct contact with
the liquid).\cite{PC} The most striking aspect of Eq.~(\ref{fpwm}) is that 
the lateral 
retention force does not depend on the size or shape of the drop, whereas 
the lateral retention force~(\ref{fpw}) is proportional to the width 
of the contact area and depends on the shape of the 
drop.\cite{FN2} Since for the experiment of Ref.~\onlinecite{TADMORPRL} the 
equilibrium contact angles are $\theta _{\rm s}= 33.0^\circ$ and 
$\theta _{\rm p}= 37.1^\circ$, Eq.~(\ref{fpwm}) yields a ratio of 1.26
for the retention forces on sessile and pendant drops,\cite{TADMORPRL} in close 
agreement with the experimental value of Eq.~(\ref{er}).

\subsection{Testing the two possible explanations}      

\begin{figure}[ht!]
\includegraphics[width=8.5cm]{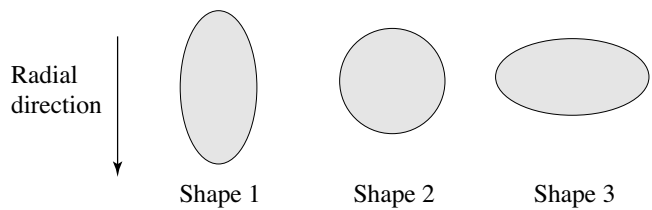}
\caption{Shapes of the contact area of 80-$\mu$L sessile drops used to compare 
Eqs.~(\ref{fpw}) and~(\ref{fpwm}).}
\label{FIG:shapes}
\end{figure}

There is no consensus yet as to whether the maximum lateral retention force
is given by Eq.~(\ref{fpw}), Eq.~(\ref{fpwm}), or some other expression. We 
have tried to compare Eqs.~(\ref{fpw}) and~(\ref{fpwm}) by preparing 
80-$\mu$L sessile drops whose contact areas have the shapes shown in 
Fig.~\ref{FIG:shapes}. We found that drops of
shape~1 started to slide first, whereas drops of shape~3 started to 
slide last.\cite{FN3} This would indicate that the correct expression for
the maximum lateral retention force is given by Eq.~(\ref{fpw}), not 
Eq.~(\ref{fpwm}). However, since for drops of shape~1 the leading edge is 
farther away from the center of rotation, whereas for drops of shape~3
it is closer to the center of rotation, one
could argue that the drops of shape~1 will move first not because
the lateral retention force is smaller but because the centrifugal force
on its leading edge is larger.\cite{PC} This loophole leaves open the 
possibility that Eq.~(\ref{fpwm}) is correct.

\section{Conclusions}
\label{sec:conclusions}

We have introduced an inexpensive and reproducible experiment
demonstrating that the retention force acting on water drops
hanging from a Plexiglas sheet is comparable to, and sometimes larger than,
the retention force on drops sitting on the surface. The
experiment consists of placing two water drops on a Plexiglas sheet,
one on top and the other on the bottom, and then rotating the Plexiglas sheet with 
an increasing angular speed. By simple visual inspection, one can see that on average the top drop starts 
to slide at about the same time as, and sometimes earlier than,
the bottom drop. In addition, the differing curvatures of 
the trails left behind by the drops also show the same result. The different
curvatures of the trails are a consequence of the Coriolis effect 
in two dimensions, which the experiment also demonstrates.

\begin{acknowledgments}
The authors would like to thank Dr.~Rafael Tadmor and Arnab Baksi for 
enlightening discussions and for lending us a camera. Special thanks 
are due to Anthony Simental, Dr.~Paul Bernazzani, Dr.~Ashwini Kucknoor, and
Emely Munda for providing some of the materials used in the experiment.
One of the authors (RM) acknowledges financial support from 
Ministerio de Ciencia e Innovaci\'on of Spain under 
project TEC2011-24492.
\end{acknowledgments}

\appendix
\section{The transverse force}
\label{sec:transf}

In this appendix, we show that the transverse force\cite{CASSIDAY}
\begin{equation}
       \vec{F}_{\rm tr}=m\left(\vec{r}\times \frac{d\vec{\Omega}}{dt} \right)
\end{equation}
can be ignored in our experiments.
Assuming that the angular acceleration $d\vec{\Omega}/dt$ of the 
platform is constant and that the object (either the magnet or the drop)
is not moving, $\vec{F}_{\rm tr}$ is a constant force that is 
perpendicular to both $\vec{r}$ and $d\vec{\Omega}/dt$. The ratio of the 
magnitudes of the centrifugal and
transverse forces can be written as
\begin{equation}
      \frac{F_{\rm cf}}{F_{\rm tr}}=\frac{mr\Omega ^2}{mr\dot{\Omega}} =
                  \frac{\Omega ^2}{\dot{\Omega}} ,
         \label{rationofm}
\end{equation}
where the dot denotes a time derivative.
This equation is valid at all instants of time. When the object
is about to slide, the angular velocity
becomes the critical angular velocity of Eq.~(\ref{nffmo}). Because the 
initial angular velocity is zero, the 
kinematic equations for constant angular acceleration tell us that
such a critical angular velocity is given by
\begin{equation}
       \Omega _{\rm critical} ^2 = 2 \dot{\Omega} \Delta \theta \, ,
           \label{kecaam}
\end{equation}
where $\Delta \theta$ is the angular displacement swept out by the
object before it starts to slide. Substitution of
Eq.~(\ref{kecaam}) into Eq.~(\ref{rationofm}) yields the ratio of the
centrifugal to the transverse force at the instant when the object is about 
to slide:
\begin{equation}
      \frac{F_{\rm cf}}{F_{\rm tr}}=2 \Delta \theta = 4\pi \cdot (\text{number of revolutions}) \, .
         \label{2rationofm}
\end{equation}

In our experiments, the magnets had to complete at least six revolutions
before they flew off, giving
\begin{equation}
      \frac{F_{\rm cf}}{F_{\rm tr}}\geq 75 \, ,  \quad 
               \text{(magnets about to slide).}
         \label{2rationofmma}
\end{equation}
Meanwhile, the drops had to complete at least ten revolutions before they
started to slide, so that
\begin{equation}
      \frac{F_{\rm cf}}{F_{\rm tr}}\geq 125  \, ,  \quad 
              \text{(drops about to slide).}
         \label{2rationofmdr}
\end{equation}
Equations~(\ref{2rationofmma}) and~(\ref{2rationofmdr}) show that at
the moment when the object is about to slide, the transverse force can 
be safely neglected compared to the centrifugal force. Thus, the force
of maximum static friction can be assumed to be equal to the value
of the centrifugal force at the instant when the object is about
to slide.

\vskip4cm

\vskip3cm

\begin{thebibliography}{99}
\bibitem{NOTE}We also noticed
that with some practice at flipping the Plexiglas sheet, the results of the
experiments improved to about 70\%--27\%--3\% for 10-$\mu$L drops and 
50\%--47\%--3\% 
for 20-$\mu$L drops.
\bibitem{FN2} We note that Eqs.~(\ref{fpw}) and~(\ref{fpwm}) are in
sharp contrast to the solid-solid frictional force, which
for rough surfaces is usually proportional to the mass (and hence the volume) 
of the object.
\bibitem{FN3} The same results were obtained when we placed the drops on
an incline.
\end{thebibliography}
\end{document}